\documentclass{article}

\usepackage[preprint]{neurips_2025}

\usepackage[utf8]{inputenc}
\usepackage[T1]{fontenc}
\usepackage{hyperref}
\usepackage{url}
\usepackage{booktabs}
\usepackage{amsfonts}
\usepackage{nicefrac}
\usepackage{microtype}
\usepackage{xcolor}
\usepackage{amsmath}
\usepackage{amssymb}
\usepackage{multirow}
\usepackage{subcaption}
\usepackage{graphicx}
\usepackage{tabularx}
\usepackage{array}
\usepackage{siunitx}
\usepackage{enumitem}
\usepackage{tikz}
\usetikzlibrary{arrows.meta, positioning, calc, fit, backgrounds, shadows.blur}
\usepackage{float}

\usepackage{adjustbox}

\newcommand{\jrrvthree}{\href{https://huggingface.co/jinaai/jina-reranker-v3}{\texttt{jina-reranker-v3}}}
\newcommand{\jrrvthreefive}{\href{https://huggingface.co/jinaai/jina-reranker-v3.5}{\texttt{jina-reranker-v3.5}}}
\newcommand{\jrrvthreefiveteacher}{\texttt{jina-reranker-v3.5-teacher}}
\newcommand{\qwenrerankersmall}{\texttt{Qwen3-Reranker-0.6B}}
\newcommand{\qwenrankerlarge}{\texttt{Qwen3-Reranker-4B}}
\newcommand{\jinaembedsmall}{\texttt{jina-embeddings-v5-text-small}}
\newcommand{\mxbaibase}{\texttt{mxbai-rerank-base-v2}}
\newcommand{\mxbailarge}{\texttt{mxbai-rerank-large-v2}}
\newcommand{\structir}{Struct-IR}

\title{\jrrvthreefive{}: An Efficient Listwise Reranker\\with Hybrid Attention and Self-Distillation}

\author{%
  Christina Nasika \quad
  Feng Wang \quad
  Antonis Krasakis \quad
  Han Xiao \\
  \\
  Jina AI \textit{by} Elastic \\
  33 New Montgomery Street, Floor 9, San Francisco, CA 94105, USA \\
  \texttt{research@jina.ai} \\
}

\begin{document}
\maketitle

\begin{abstract}
Listwise rerankers are the discriminative core of agentic retrieval pipelines, yet production deployment demands efficiency, domain robustness, and fluency on semi-structured data at the same time. We present \jrrvthreefive{}, a 0.6B-parameter listwise reranker that meets these demands together without sacrificing the cross-document comparison that makes its predecessor \jrrvthree{} effective. \jrrvthreefive{} keeps the last-but-not-late (LBNL) interaction of \jrrvthree{} and reworks it along three axes. It replaces uniform global attention with a hybrid schedule of three sliding-window layers followed by two global layers, pinning the terminal layer to global as LBNL readout requires. It trains on a curated multi-domain mixture that spans legal, medical, financial, multilingual, and structured retrieval. It transfers quality through a three-stage self-distillation recipe in which a full-attention teacher sets an upper bound that a sparse-attention student then recovers under a staged adaptation protocol. \jrrvthreefive{} reaches 63.20 nDCG@10 on BEIR, matching a 4B model at roughly $7\times$ fewer parameters, and improves over \jrrvthree{} on MIRACL and RTEB as well. Its largest gains come on semi-structured retrieval, where it lifts nDCG@10 by 9.6 points over \jrrvthree{} and leads all rerankers of comparable size. The hybrid schedule further cuts listwise inference latency by up to $1.56\times$. We release the model weights on Hugging Face under a non-commercial license.
\end{abstract}

\section{Introduction}
\label{sec:introduction}

Neural rerankers sit at the last stage of information retrieval pipelines. A first-stage retriever proposes a short candidate list, and the reranker reorders it under tight latency and memory budgets. As retrieval moves inside agentic systems that issue many queries over long-horizon tasks, the reranker is called repeatedly and its cost and reliability compound. A production reranker must therefore meet several demands that are easy to satisfy individually but hard to meet together.
The first demand is efficiency. Listwise rerankers score many candidates in one forward pass, so full self-attention grows quadratically with the candidate list and inflates the key-value cache. The second is domain robustness. Legal, medical, financial, and e-commerce corpora diverge sharply from the web-scraped text that dominates standard training mixtures, and a model with top BEIR~\citep{thakur2021beir} scores can still fail under domain-specific production conditions. The third is semi-structured data understanding. Much real content arrives as JSON records, data tables, and key-value fields, where relevance depends on matching a query to specific fields such as entities, dates, numeric ranges, and categorical attributes rather than on surface lexical overlap~\citep{stark,ssrb}. The fourth is multilingual coverage, since production retrieval routinely spans mixed-language data~\citep{zhang2022miracl}.
\jrrvthree{}~\citep{wang2025jinarerankerv3} advanced listwise reranking through its last-but-not-late (LBNL) interaction. The query and all candidates form a single causal sequence, self-attention runs over this joint context, and a lightweight MLP then projects the contextual embeddings of special delimiter tokens into an output space where query and document vectors are compared by cosine similarity. Joint encoding gives LBNL in-context cross-document comparison, which late-interaction models such as ColBERT~\cite{khattab2020colbert} lack because they encode passages independently. The 0.6B \jrrvthree{} reached 62.10 nDCG@10 on BEIR, the state of the art at its scale.

That model addresses only part of the deployment picture. It runs full global attention at every layer, trains mainly on general-domain retrieval, and sees little field-constrained ranking on semi-structured data. \jrrvthreefive{} is a 0.6B LBNL successor that targets all four demands at once, improving efficiency, domain coverage, semi-structured understanding, and multilingual transfer without giving up listwise quality.

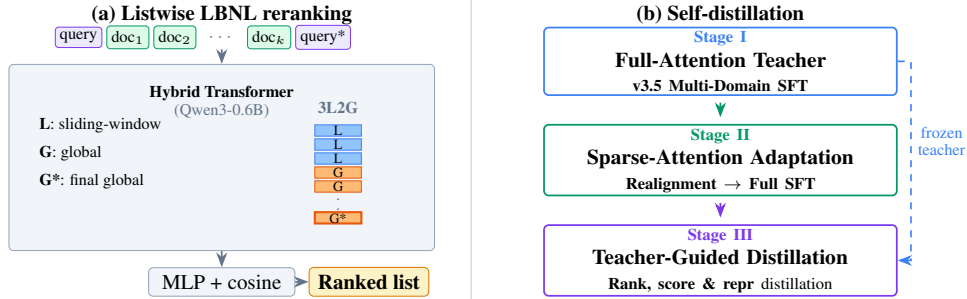
\begin{figure}[H]
  \centering
  \vspace{-4mm}
  \setlength{\abovecaptionskip}{4pt}%
  \setlength{\belowcaptionskip}{0pt}%
  \resizebox{0.95\textwidth}{!}{
\begin{tikzpicture}[
  x=1mm, y=1mm,
  font=\footnotesize,
  >=Stealth,
  arr/.style={->, line width=0.75pt, draw=fgmid},
  arrdash/.style={line width=0.70pt, draw=stage1, dashed},
  tokq/.style={draw=qedge, fill=qfill, rounded corners=1.5pt, inner sep=2pt, font=\scriptsize},
  tokd/.style={draw=dedge, fill=dfill, rounded corners=1.5pt, inner sep=2pt, font=\scriptsize},
  block/.style={draw=boxedge, fill=boxfill, rounded corners=2pt, align=center, inner sep=3pt},
  stage/.style n args=1{draw=#1, fill=white, rounded corners=2pt, align=center,
    inner sep=1.5pt, line width=0.75pt, text width=52mm},
  layer/.style={minimum width=7mm, minimum height=1.7mm, inner sep=0pt},
  loc/.style  ={layer, fill=localfill,  draw=localedge,  line width=0.3pt,
    font=\tiny, text=black, align=center},
  glob/.style ={layer, fill=globalfill, draw=globaledge, line width=0.3pt,
    font=\tiny, text=black, align=center},
  gstar/.style={layer, fill=globalfill, draw=globaledge, line width=0.9pt,
    font=\tiny, text=black, align=center},
]
  \definecolor{fgmid}{HTML}{64748B}
  \definecolor{boxfill}{HTML}{F1F5F9}
  \definecolor{boxedge}{HTML}{94A3B8}
  \definecolor{qfill}{HTML}{EDE9FE}
  \definecolor{qedge}{HTML}{7C3AED}
  \definecolor{dfill}{HTML}{DCFCE7}
  \definecolor{dedge}{HTML}{059669}
  \definecolor{localfill}{HTML}{93C5FD}
  \definecolor{localedge}{HTML}{3B82F6}
  \definecolor{globalfill}{HTML}{FDBA74}
  \definecolor{globaledge}{HTML}{EA580C}
  \definecolor{stage1}{HTML}{3B82F6}
  \definecolor{stage2}{HTML}{059669}
  \definecolor{stage3}{HTML}{7C3AED}
  \definecolor{scorefill}{HTML}{FEF3C7}
  \definecolor{scoreedge}{HTML}{D97706}

  \pgfmathsetmacro{\halfW}{75}
  \pgfmathsetmacro{\cxA}{37.5}
  \pgfmathsetmacro{\cxB}{112.5}

  \useasboundingbox (0, -48) rectangle (150, 4);

  \node[anchor=north, font=\small\bfseries] at ({\cxA},  0) {(a) Listwise LBNL reranking};
  \node[anchor=north, font=\small\bfseries] at ({\cxB},  0) {(b) Self-distillation};

  \node[font=\scriptsize, text=fgmid] (ell) at ({\cxA}, -6) {$\cdots$};
  \node[tokd, left=0.6mm of ell]  (d2) {doc$_2$};
  \node[tokd, left=0.6mm of d2]   (d1) {doc$_1$};
  \node[tokq, left=0.6mm of d1]   (q)  {query};
  \node[tokd, right=0.6mm of ell] (dk) {doc$_k$};
  \node[tokq, right=0.6mm of dk]  (qs) {query*};

  \begin{scope}[on background layer]
    \node[block, minimum width=64mm, minimum height=28mm,
          anchor=north, inner sep=0pt] (bb) at ({\cxA}, -10) {};
  \end{scope}

  \node[anchor=north, font=\scriptsize\bfseries] at ($(bb.north)+(0,-1.8)$)
    {\strut Hybrid Transformer};
  \node[anchor=north, font=\scriptsize, text=fgmid]
    at ($(bb.north)+(0,-4.5)$) {\strut (Qwen3-0.6B)};

  \node[anchor=north west, align=left, inner sep=0pt] at ({10}, -18.0) {%
    {\scriptsize \textbf{L}:\ sliding-window}\\[2pt]
    {\scriptsize \textbf{G}:\ global}\\[2pt]
    {\scriptsize \textbf{G*}: final global}%
  };

  \node[anchor=north, font=\scriptsize\bfseries, text=fgmid] (patlab)
    at ({55}, -14.5) {3L2G};
  \node[loc,  below=0.4mm of patlab, anchor=north] (lay1) {L};
  \node[loc,  below=0.3mm of lay1,   anchor=north] (lay2) {L};
  \node[loc,  below=0.3mm of lay2,   anchor=north] (lay3) {L};
  \node[glob, below=0.3mm of lay3,   anchor=north] (lay4) {G};
  \node[glob, below=0.3mm of lay4,   anchor=north] (lay5) {G};
  \node[gstar,below=2.8mm of lay5.south,  anchor=north] (layN) {G*};
  \node[font=\scriptsize, text=fgmid, anchor=center, inner sep=0pt] (vdots) 
    at ($(lay5.south)!0.5!(layN.north)$) {$\vdots$};

  \node[block, anchor=south, minimum width=22mm] (head)
    at ({\cxA}, -45) {MLP + cosine};
  \node[block, anchor=south west, minimum width=18mm,
        draw=scoreedge, fill=scorefill] (out)
    at ($(head.south east)+(2,0)$) {\textbf{Ranked list}};

  \draw[arr] (ell.south) -- (bb.north);   
  \draw[arr] (bb.south)  -- (head.north); 
  \draw[arr] (head.east) -- (out.west);   

  \coordinate (bAxis) at ({\cxB}, 0);
  \coordinate (bTop) at (bAxis |- q.north);
  \coordinate (bBot) at (bAxis |- out.south);

  \node[stage=stage1, anchor=north] (s1) at (bTop) {%
    \textcolor{stage1}{\scriptsize\bfseries Stage I}\\[0.5pt]
    \textbf{Full-Attention Teacher}\\[0.5pt]
    {\scriptsize \textbf{v3.5 Multi-Domain SFT}}%
  };
  \node[stage=stage3, anchor=south] (s3) at (bBot) {%
    \textcolor{stage3}{\scriptsize\bfseries Stage III}\\[0.5pt]
    \textbf{Teacher-Guided Distillation}\\[0.5pt]
    {\scriptsize \textbf{Rank, score \& repr} distillation}%
  };
  \node[stage=stage2, anchor=center] (s2) at ($(s1.south)!0.5!(s3.north)$) {%
    \textcolor{stage2}{\scriptsize\bfseries Stage II}\\[0.5pt]
    \textbf{Sparse-Attention Adaptation}\\[0.5pt]
    {\scriptsize \textbf{Realignment} $\rightarrow$ \textbf{Full SFT}}%
  };

  \draw[arr, draw=stage2] ([yshift=-1mm]s1.south) -- ([yshift=1mm]s2.north);
  \draw[arr, draw=stage3] ([yshift=-1mm]s2.south) -- ([yshift=1mm]s3.north);

  \path (s1.east) -- ++(2, 0) coordinate (ftx);
  \draw[arrdash, ->] (s1.east) -- (ftx) -- (ftx |- s3.east) -- (s3.east);
  \node[anchor=west, align=center, font=\scriptsize, text=stage1,
        inner sep=0pt, xshift=1mm]
    at ($(ftx)!0.40!(ftx |- s3.east)$) {frozen\\[-1pt]teacher};

  \draw[fgmid!40, line width=0.4pt] ({\halfW}, 0.5) -- ({\halfW}, -47);

\end{tikzpicture}}%
  \vspace{-3mm}
  \caption{Overview of \jrrvthreefive{}.
  \textbf{(a)}~LBNL listwise encoding with a 3L2G hybrid backbone and pinned terminal global layer~(G*).
  \textbf{(b)}~Three-stage self-distillation from a full-attention teacher to a sparse-attention student.}
  \label{fig:overview}
\end{figure}

Figure~\ref{fig:overview} summarizes the model architecture. Three coordinated changes distinguish \jrrvthreefive{} from its predecessor.

\begin{enumerate}[label=\roman*]
    \item A hybrid 3L2G attention schedule that repeats three sliding-window layers followed by two global layers, keeping the terminal global layer that LBNL interaction requires.
    \item A curated multi-domain training mixture that spans legal, medical, financial, multilingual, and semi-structured retrieval.
    \item A three-stage self-distillation recipe that starts from a similarly-sized full-attention teacher, adapts it under a sparse-attention mask, and distills the result into the hybrid-attention student.
\end{enumerate}

The distillation step is unusual in that we do not distill from a larger model into a smaller one. Teacher and student have the same size and differ only in attention pattern. The teacher is trained under full attention with its quadratic cost, and its behavior is distilled into a student that runs the more efficient 3L2G schedule. Decoupling the two problems this way spares the student from relearning its routing and matching the teacher's output at the same time.

\jrrvthreefive{}'s weights are available on Hugging Face for non-commercial use.

\section{Related Work}
\label{sec:related}

Neural reranking builds on the learning-to-rank tradition, whose pointwise, pairwise, and listwise objectives frame how candidate lists are scored~\cite{burges2005learning,li2007mcrank}. Cross-encoders score query and document pairs jointly but need a separate forward pass per pair~\cite{nogueira2019passage,dejean2024thorough}. ColBERT-style late-interaction models instead decouple encoding from matching and cache large multivector representations of queries and documents independently, at the cost of losing cross-document comparison and query-conditioned context~\cite{khattab2020colbert,liu2024analysis,jha2024jina}. LBNL-interaction reranking combines joint encoding with cosine scoring, recovering cross-document and query awareness at moderate computational cost~\cite{wang2025jinarerankerv3}.

Efficient long-context transformers offer the mechanisms we build on to cut that cost. Sliding-window and local-global attention reduce the quadratic complexity of self-attention~\cite{beltagy2020longformer,jiang2023mistral}, and Gemma~3 shows that interleaving local and global layers in a fixed ratio preserves long-context language modeling quality~\cite{gemmateam2025gemma3}. Our setting differs from these generative models in two ways that shape the design. A reranker has no autoregressive output budget to amortize its context over, and LBNL interaction requires the final layer to remain global so the trailing query embedding can observe the entire candidate list. This readout constraint is absent in generative language modeling and distinguishes our hybrid backbone from prior local-global systems.

Knowledge distillation under capacity mismatch is the second body of work we draw on. Distillation is a standard route to model compression~\cite{hinton2015distilling,agarwal2023onpolicy}, but naive direct transfer underperforms when student and teacher are structurally mismatched~\cite{mirzadeh2020improved}. Prior work bridges size gaps with an intermediate teacher assistant~\cite{mirzadeh2020improved}. Our mismatch is not size but attention pattern, and our three-stage recipe plays the assistant role with the student itself at an intermediate training phase, adapting under the sparse mask before matching teacher outputs.

Domain-specific and field-constrained retrieval motivates the training mixture. The RTEB benchmark suite shows that unspecialized rerankers degrade on legal, medical, and financial text~\cite{rteb2025}, and they also fall short on the STARK~\cite{stark} and Struct-IR~\cite{ssrb} benchmarks over product catalogs and knowledge graphs. These benchmarks are usually treated as evaluation targets. We instead use them to diagnose failure modes and curate training data that addresses the shortcomings they expose.

\section{Method}
\label{sec:method}

\subsection{Hybrid Attention for LBNL Reranking}
\label{sec:hybrid-attention}

We briefly recap the \jrrvthree{} interface that \jrrvthreefive{} inherits, with full details in~\cite{wang2025jinarerankerv3}.
Given query $q$ and candidates $\{d_i\}_{i=1}^{k}$, a listwise prompt concatenates all passages with delimiter tokens and places the query in a trailing block.
Causal self-attention over the full sequence produces contextual embeddings $\tilde{\mathbf{q}}$ and $\{\tilde{\mathbf{d}}_i\}$ at the special token positions; these are projected by a two-layer MLP $P_\phi$ into a 512-dimensional space and scored by cosine similarity:
\begin{equation}
  s_i = \cos\bigl(P_\phi(\tilde{\mathbf{q}}),\; P_\phi(\tilde{\mathbf{d}}_i)\bigr).
\end{equation}
The model is trained with an InfoNCE listwise ranking loss and auxiliary dispersive, dual-matching, and similarity objectives~\cite{wang2025jinarerankerv3,oord2019representation,huang2024cosent,wang2024contrastive}.

The pinned terminal global layer is the central architectural constraint of \jrrvthreefive{}. Full self-attention at all 28 layers dominates compute when candidate lists are long, and the natural response is to replace most layers with sliding-window attention~\cite{beltagy2020longformer,jiang2023mistral}. The LBNL readout forbids a uniform switch. The query embedding token sits at the end of the concatenated sequence and must attend back to the first candidate to form a cross-document-aware representation, and a finite window $w$ severs that long-range dependency. We therefore keep the final layer global at all times, so the query and document embedding tokens observe the entire candidate context at the point of extraction. We call this the pinned terminal global layer, shown as G* in Figure~\ref{fig:overview}(a). Replacing it with a sliding window severely degraded listwise ranking because the trailing query embedding could no longer observe early candidates, whereas keeping only the final layer global preserved joint encoding without a fully global stack.

Beyond the terminal layer, we searched over local-to-global ratios in the remaining 27 layers, comparing 1L1G, 1L2G, 3L2G, and 5L1G schedules on held-out development sets. The 3L2G schedule repeats three consecutive sliding-window layers followed by two global layers, and it gave the best accuracy and efficiency trade-off. Under 3L2G the 28-layer backbone contains 17 local and 11 global layers with a sliding-window span of $w=1024$ tokens. Local layers reduce attention complexity from $\mathcal{O}(L^2)$ to $\mathcal{O}(L \cdot w)$ per layer, while the global layers placed every three steps periodically refresh long-range cross-candidate signals throughout the network depth. Distributing the global layers uniformly preserves information flow at every depth, which we found preferable to front-loading or back-loading them. Denser schedules such as 1L1G and 1L2G gave little throughput benefit, and the more aggressive 5L1G trended worse on complex multi-document tasks, leaving 3L2G as the best point for latency and KV-cache savings.

\subsection{Multi-Domain Training Mixture}
\label{sec:data}

Our curation follows a failure-mode-first principle rather than a data-volume one. The \jrrvthree{} mixture already covers general retrieval well but underrepresents specialized domains, so each new shard is built to cover the retrieval patterns that general models fail on, identified through error analysis on the RTEB and STARK development sets. To prevent shortcut learning we source hard negatives from several retrievers, including BM25, Jina, BGE, GTE, E5, and ColBERT. For domains with sparse relevance labels we synthesize queries with an LLM conditioned on passage content and domain-specific relevance rubrics.

The legal shard combines EUR-Lex multilingual, CLERC, AILA, Canadian case law, Swiss case summarization, and EuroVoc with LLM-generated user queries and jurisdiction-specific hard negatives. Legal text is citation-dense and long, so we oversample these shards to compensate for their higher annotation complexity. The medical shard draws on MIRIAD-style dialogue, biomedical citation retrieval, PubMed GPL negatives, and CMedQA-style Chinese medical QA from the BGE mixture, targeting clinical wording and entity-heavy passages. The finance shard uses FiQA investment forum QA, financial QA benchmarks, and table-aware passages from the RTEB finance tasks, prioritizing numeric claims, regulatory language, and structured financial records.

\begin{figure*}[t]
  \centering
  \resizebox{\textwidth}{!}{
\begin{tikzpicture}[
  font=\small,
  >={Stealth[round,length=2.0mm,width=1.6mm,sep=0.8pt]},
  card/.style={rectangle, rounded corners=5pt, line width=1pt, draw=#1,
    text=cInk, align=center, minimum height=10mm, inner sep=4.5pt},
  card/.default=cProc,
  proc/.style ={card=cProc, fill=cProcBg, text width=52mm},
  data/.style ={card=cData, fill=cDataBg, text width=52mm},
  llm/.style  ={card=cLLM,  fill=cLLMBg,  text width=52mm},
  retr/.style ={card=cRetr, fill=cRetrBg, text width=52mm},
  outbox/.style ={card=cOut, fill=cOutBg, text width=52mm},
  flow/.style={->, line width=1.05pt, draw=cInk!70},
  loopflow/.style={->, line width=1.05pt, draw=cRetr!85!black, densely dashdotted,
    >={Stealth[round,length=1.8mm,width=1.5mm,sep=2.8pt]}},
  title/.style={font=\bfseries\small, text=cInk},
  doc/.style ={rectangle, rounded corners=5pt, draw=#1, line width=1.1pt, fill=white,
    text=cInk, align=left, inner sep=5.5pt, text width=42mm, font=\scriptsize},
  doc/.default=cProc,
  bubble/.style={rectangle, rounded corners=8pt, draw=cLLM, line width=1.1pt,
    fill=cLLMBg, text=cInk, align=center, inner sep=6pt, text width=74mm},
  pill/.style ={rounded corners=4pt, draw=#1!70, line width=0.7pt, fill=#1!12,
    text=#1!72!black, font=\scriptsize\bfseries, inner xsep=4pt, inner ysep=2pt},
  pill/.default=cProc,
  cornerok/.style={circle, fill=cRetr, text=white, font=\scriptsize\bfseries,
    inner sep=0.3pt, minimum size=4.2mm, draw=white, line width=0.6pt},
  cornerno/.style={circle, fill=cRej, text=white, font=\scriptsize\bfseries,
    inner sep=0.3pt, minimum size=4.2mm, draw=white, line width=0.6pt},
  spot/.style={font=\bfseries\small, text=cInk, align=center},
  callout/.style={rectangle, rounded corners=7pt, draw=cRetr, line width=1.1pt,
    fill=cRetrBg, text=cInk, align=left, inner sep=5pt, text width=28mm,
    font=\scriptsize},
]
  \definecolor{cInk}{HTML}{2B2D42}
  \definecolor{cData}{HTML}{F4A100}  \definecolor{cDataBg}{HTML}{FFF4D9}
  \definecolor{cProc}{HTML}{3A86FF}  \definecolor{cProcBg}{HTML}{E7F0FF}
  \definecolor{cLLM}{HTML}{8338EC}   \definecolor{cLLMBg}{HTML}{F3EAFF}
  \definecolor{cRetr}{HTML}{06A77D}  \definecolor{cRetrBg}{HTML}{DEF6EF}
  \definecolor{cOut}{HTML}{2A9D8F}   \definecolor{cOutBg}{HTML}{DAF3EF}
  \definecolor{cRej}{HTML}{E5383B}
  \definecolor{cPanel}{HTML}{F7F8FC} \definecolor{cPanelLine}{HTML}{DDE3F0}

  \node[data] (s1)
     {\textbf{Anchor record}\\[1pt]
      {\scriptsize JSON object with typed fields}};
  \node[llm, below=5.5mm of s1] (s2)
     {\textbf{(i)~Constraints $\rightarrow$ query}\\[1pt]
      {\scriptsize Sample typed field constraints;\\[-0.5pt]
       an LLM paraphrases them into a query}};
  \node[proc, below=5.5mm of s2] (s3)
     {\textbf{(ii)~Perturb $\rightarrow$ hard negative}\\[1pt]
      {\scriptsize Change 1--2 constrained fields to form\\[-0.5pt]
       a near-duplicate hard negative}};
  \node[retr, below=5.5mm of s3] (s4)
     {\textbf{(iii)~Dense-mine candidates}\\[1pt]
      {\scriptsize Retrieve additional near neighbors}};
  \node[llm, below=5.5mm of s4] (s5)
     {\textbf{(iv)~LLM-judge the list}\\[1pt]
      {\scriptsize Promote true matches, refine over-broad\\[-0.5pt]
       queries, discard ambiguous cases}};
  \node[outbox, below=5.5mm of s5] (s6)
     {\textbf{Training instance}\\[1pt]
      {\scriptsize query $+$ \textcolor{cRetr}{\checkmark}\,positives $+$ \textcolor{cRej}{$\boldsymbol\times$}\,hard negatives}};

  \foreach \a/\b in {s1/s2, s2/s3, s3/s4, s4/s5, s5/s6}
    \draw[flow] (\a) -- (\b);

  \node[callout, left=16mm of s5] (loop)
     {\textbf{\textcolor{cRetr}{Feedback}}\\[2pt]
      re-enter at~(i)\\[1pt]
      after promote/refine/discard};

  \node[bubble, right=24mm of s2] (qb)
     {\emph{``trail-running shoes \textbf{under \$150}, \textbf{rated 4.5+}, \textbf{released since 2022}''}};
  \node[spot, above=3.5mm of qb, text width=80mm] (sptitle)
     {Hard negative keeps the same surface form\\[-1pt]
      but violates one (or two) constraints};

  \node[pill=cLLM, below=3mm of qb.south, xshift=-26mm] (p1) {price $\le$ \$150};
  \node[pill=cLLM, right=2mm of p1] (p2) {rating $\ge$ 4.5};
  \node[pill=cLLM, right=2mm of p2] (p3) {year $\ge$ 2022};

  \node[doc=cRetr, below left=14mm and 3mm of qb.south] (pos)
     {\textbf{\textcolor{cRetr}{Positive}} (meets all constraints)\\[2.5pt]
      product:\ trail-running shoe\\
      price:\ \$129\\
      rating:\ 4.7 / 5\\
      released:\ 2023};
  \node[doc=cRej, below right=14mm and 3mm of qb.south] (neg)
     {\textbf{\textcolor{cRej}{Hard negative}} (one field perturbed)\\[2.5pt]
      product:\ trail-running shoe\\
      \textcolor{cRej}{\textbf{price:\ \$210}}\ {\scriptsize\textcolor{cRej}{($>$ \$150)}}\\
      rating:\ 4.7 / 5\\
      released:\ 2023};

  \node[cornerok] at (pos.north east) {\checkmark};
  \node[cornerno] at (neg.north east) {$\boldsymbol\times$};

  \draw[flow] (pos.north |- p1.south) ++(0,-2mm) -- (pos.north);
  \draw[flow] (neg.north |- p1.south) ++(0,-2mm) -- (neg.north);

  \begin{scope}[on background layer]
    \node[fit=(s1)(s6)(loop)(sptitle)(pos)(neg), rounded corners=10pt,
          fill=cPanel, draw=cPanelLine, line width=0.9pt, inner sep=8mm] (panel) {};
    \node[title, anchor=north west] at ([shift={(3.5mm,-2.5mm)}]panel.north west)
      {\structir{} synthetic supervision};
    \node[fit=(sptitle)(qb)(pos)(neg)(p1)(p3), rounded corners=8pt,
          fill=white, draw=cPanelLine, line width=0.9pt, inner sep=4.5mm] {};
  \end{scope}

  \coordinate (aOut) at ([xshift=-2.2mm]s5.west);
  \coordinate (aIn)  at ([xshift=2.2mm]loop.east);
  \coordinate (bOut) at ([yshift=2.2mm]loop.north);
  \coordinate (bUp)  at ([yshift=7.5mm]loop.north);
  \coordinate (bIn)  at ([xshift=-2.2mm]s2.west);
  \draw[line width=1.05pt, draw=cRetr!85!black, densely dashdotted]
    (s5.west) -- (aOut)
    (aIn) -- (loop.east)
    (loop.north) -- (bOut)
    (bIn) -- (s2.west);
  \fill[cRetr!85!black] (aOut) circle (0.6pt)
                        (aIn)  circle (0.6pt)
                        (bOut) circle (0.6pt)
                        (bIn)  circle (0.6pt);
  \draw[loopflow] (aOut) -- (aIn);
  \draw[loopflow] (bOut) -- (bUp) -- (bUp |- bIn) -- (bIn);
\end{tikzpicture}}%
  \caption{Constraint-heavy synthetic data for \structir{}.
  \emph{Left:} (i)~constraints$\rightarrow$query, (ii)~perturb hard negative, (iii)~dense-mine, (iv)~LLM-judge with a promote/refine/discard feedback loop.
  \emph{Right:} near-duplicate positive vs.\ hard negative differing by one field.}
  \label{fig:structir-pipeline}
\end{figure*}
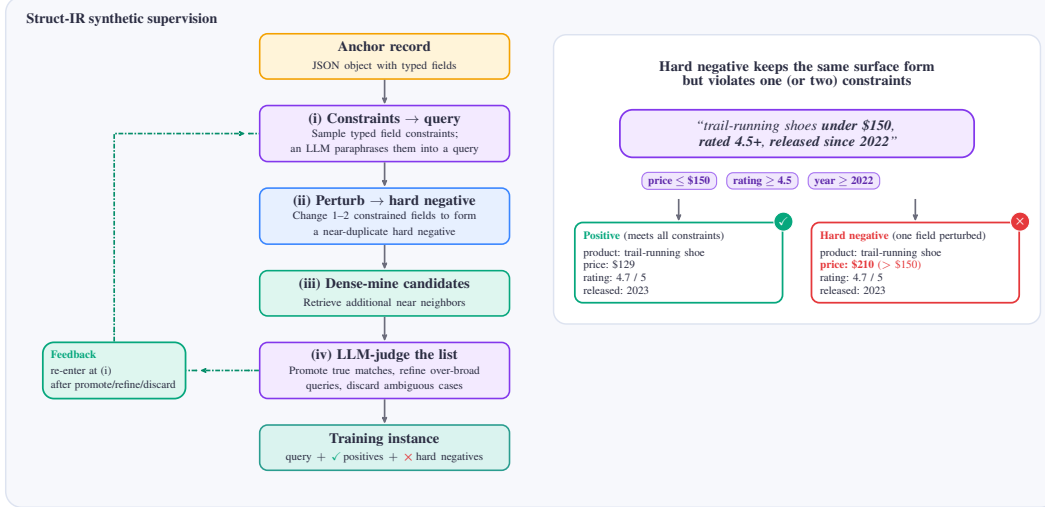

Structured data receives the most attention because it falls outside the free-text distribution of standard benchmarks, so we treat it as a dedicated shard. STARK entity attributes~\citep{stark}, serialized as flat records rather than relational graphs, and Struct-IR~\citep{ssrb} supply field-level ranking over heterogeneous schemas. Table sources from Open-WikiTable~\citep{openwikitable}, NQ-Tables~\citep{nqtables}, OTT-QA~\citep{ottqa}, ESCI~\citep{esci}, TabFact~\citep{tabfact}, TAT-QA~\citep{tatqa}, SQA~\citep{sqa}, WikiTableQuestions~\citep{wtq}, and HybridQA~\citep{hybridqa} add cell-level supervision. Relevance here hinges on equality, numeric and date bounds, list membership, and logical combinations across fields rather than bag-of-words overlap. Early runs lagged most on this slice, so we assign it the highest sampling weight and synthesize additional constraint-heavy pairs, illustrated in Figure~\ref{fig:structir-pipeline}. The pipeline samples typed constraints from an anchor record and paraphrases them into a query, perturbs one or two constrained fields to form a near-duplicate hard negative, dense-mines further candidates, and finally has an LLM judge the list to promote true matches, refine over-broad queries, and discard ambiguous cases.

Multilingual coverage extends beyond the MIRACL and mMARCO shards used for \jrrvthree{}. We add WebFAQ contexts in more than 50 languages, SWIMIR cross-lingual hard negatives, Ruri-v3 Japanese reranker pairs, and refreshed MIRACL hard-negative sets reranked with \jinaembedsmall{}.

\subsection{Three-Stage Self-Distillation}
\label{sec:training}

The three stages address a single challenge. The full-attention teacher and the sparse-attention student have different information-routing capacities, so forcing a direct attention-mask switch while also matching the teacher's output distributions makes the student underperform on both ranking quality and representation alignment. Our recipe decouples these two pressures, as shown in Figure~\ref{fig:overview}(b).

Stage~I trains a full-attention teacher. Starting from the public \jrrvthree{} checkpoint, we fully fine-tune \jrrvthreefiveteacher{} without any sliding-window restrictions, using the complete data mixture from Section~\ref{sec:data} with domain-tuned sampling rates. The teacher establishes a quality upper bound, since its listwise scores under full attention on the v3.5 data distribution represent the best achievable performance at this parameter budget.

Stage~II adapts a student to sparse attention. We initialize the student from the Stage~I weights and activate the 3L2G attention pattern, then adapt in two sub-stages. The first sub-stage, attention realignment, trains only the attention projection matrices while the MLP, embedding, and head layers stay frozen, which teaches the sparse masks to route information efficiently without disrupting the representations learned under full attention. The second sub-stage unfreezes all parameters, optionally with LoRA on selected linear layers, and lets the student realign its intermediate representations to the new attention geometry. After Stage~II the student is already deployable and faster than the teacher, but development evaluations still show a consistent quality gap on BEIR, RTEB-legal, and MIRACL.

Stage~III closes that gap through teacher-guided distillation. With \jrrvthreefiveteacher{} frozen, we continue training the Stage~II student under a multi-level objective that aligns it to the teacher across output scores and intermediate representation geometries, combined with self-supervised auxiliary regularization:
\begin{equation}
  \mathcal{L} = \mathcal{L}_{\mathrm{rank}} + \beta_{1}\,\mathcal{L}_{\mathrm{score}} + \beta_{2}\,\mathcal{L}_{\mathrm{state}} + \beta_{3}\,\mathcal{L}_{\mathrm{embed}} + \alpha_{1}\,\mathcal{L}_{\mathrm{sim}} + \alpha_{2}\,\mathcal{L}_{\mathrm{disp}}
\end{equation}
where each term corresponds precisely to a component in our distillation engine:
\begin{itemize}[leftmargin=*]
  \item \textbf{Rank-Level Alignment ($\mathcal{L}_{\mathrm{rank}}$):}
  A listwise Kullback--Leibler (KL) divergence aligns the softmax-normalized probability distributions over candidate document batches, preserving relative ranking structures invariant to score scale:
  \begin{equation}
    \mathcal{L}_{\mathrm{rank}} = \mathrm{KL}\bigl(\mathrm{Softmax}(\mathbf{s}_{\mathrm{tch}} / \tau) \parallel \mathrm{Softmax}(\mathbf{s}_{\mathrm{std}} / \tau)\bigr),
  \end{equation}
  scaled by a distillation temperature $\tau = 0.25$.
  \item \textbf{Score-Level Alignment ($\mathcal{L}_{\mathrm{score}}$):}
  An MSE loss preserves absolute score scaling and classification margins across predicted candidate scores:
  \begin{equation}
    \mathcal{L}_{\mathrm{score}} = \frac{1}{K} \sum_{i=1}^{K} (s_{\mathrm{std}, i} - s_{\mathrm{tch}, i})^2,
  \end{equation}
  weighted by $\beta_{1} = 0.65$.
  \item \textbf{State-Level Alignment ($\mathcal{L}_{\mathrm{state}}$):}
  An MSE loss over the sequence of last-layer hidden states $\mathbf{h}_t \in \mathbb{R}^H$:
  \begin{equation}
    \mathcal{L}_{\mathrm{state}} = \frac{1}{T \cdot H} \sum_{t=1}^{T} \lVert \mathbf{h}_{\mathrm{std}, t} - \mathbf{h}_{\mathrm{tch}, t} \rVert_2^2,
  \end{equation}
  weighted by $\beta_{2} = 0.65$.
  \item \textbf{Embedding Cosine Alignment ($\mathcal{L}_{\mathrm{embed}}$):}
  A cosine distance loss over the projected embeddings $\mathbf{e}_j$ (including query and document tokens):
  \begin{equation}
    \mathcal{L}_{\mathrm{embed}} = 1 - \frac{1}{1 + K} \sum_{j=1}^{1+K} \cos(\mathbf{e}_{\mathrm{std}, j}, \mathbf{e}_{\mathrm{tch}, j}),
  \end{equation}
  weighted by $\beta_{3} = 0.65$.
  \item \textbf{Self-Supervised \& Auxiliary Losses ($\mathcal{L}_{\mathrm{sim}}$, $\mathcal{L}_{\mathrm{disp}}$):}
  To maintain robust embedding spaces under sparse attention, we integrate in-context contrastive chunk similarity classification $\mathcal{L}_{\mathrm{sim}}$ (weighted by $\alpha_{1} = 0.65$) and token embedding dispersion regularization $\mathcal{L}_{\mathrm{disp}}$ (weighted by $\alpha_{2} = 0.25$).
\end{itemize}

This purely teacher-guided multi-level formulation updates the student parameters to mimic the teacher's entire representation hierarchy, and we cache teacher forward passes where possible to reduce per-step overhead. We train several domain-specific distillation runs independently, covering a default mixture together with BGE-heavy, legal-heavy, and long-document MLDR variants, and merge them by linear interpolation with weights tuned on development RTEB and BEIR nano subsets. Stage~III recovers the performance gap between the Stage~II student and the Stage~I teacher, transferring teacher knowledge across mismatched attention geometries. The ordering matters. Stage~II on its own, activating 3L2G without teacher matching, leaves a noticeable gap to the teacher because the student must change its routing under a weaker mask before it can safely match teacher scores and states. Stage~III then recovers most of that gap, which shows that full-attention capacity can transfer into a sparse student once the geometry has adapted.

All stages use FlashAttention-2, bf16 precision, listwise batches of up to 50 documents, and 30 to 45 hard negatives per query. Stage~I uses a learning rate of $6\times10^{-6}$ for up to 25k steps. Stage~II follows a two-phase schedule with an attention-only warmup at $5\times10^{-5}$ and then full-parameter tuning at $6\times10^{-6}$, for up to 15k plus 15k steps. Stage~III uses $5\times10^{-5}$ for 5k to 25k steps depending on the domain shard. The distillation temperature $\tau$ ranges from 0.05 to 0.25 across stages. We apply random document-ID and position-ID perturbations throughout to improve listwise robustness~\cite{wang2025jinarerankerv3}.

\section{Evaluation}
\label{sec:evaluation}

\subsection{Experimental Setup}
\label{sec:setup}

We evaluate across general, multilingual, domain-specific, and structured retrieval regimes.
BEIR~\cite{thakur2021beir} tests zero-shot English generalization across 13 heterogeneous datasets; MIRACL~\cite{zhang2022miracl} covers 18 languages; RTEB~\cite{rteb2025} targets professional-domain corpora in legal, finance, programming, and medicine; and STARK~\cite{stark} and Struct-IR~\cite{ssrb} evaluate structured and field-constrained retrieval.

Unless noted otherwise, top-100 candidates are retrieved with \jinaembedsmall{}~\cite{akram2026jinaembeddingsv5texttasktargetedembeddingdistillation} and reranked listwise.
Task-specific LoRA adapters are enabled for ArguAna and Quora.
All baselines are re-evaluated under this unified MTEB~v2 pipeline; numbers may differ slightly from prior reports.
Baselines include \jrrvthree{}, \qwenrerankersmall{}, \qwenrankerlarge{}, \mxbaibase{}, and \mxbailarge{}~\cite{shakir2025mxbairerankv2,yang2025qwen3embedding}.
We report nDCG@10 everywhere.
Efficiency measurements use a single NVIDIA A100, batch size~1, and top-100 listwise reranking with FlashAttention-2 (Section~\ref{sec:efficiency}).

\subsection{Overall Results}
\label{sec:overall}

\jrrvthreefive{} improves on its predecessor across every benchmark family while keeping the same parameter count and inference interface. Table~\ref{tab:overall} reports the macro results. On BEIR it reaches 63.20 nDCG@10 against 62.10 for \jrrvthree{}, and it also edges past the 4B \qwenrankerlarge{} at 62.28 with roughly $7\times$ fewer parameters, with per-dataset scores in Appendix~\ref{app:beir}. On MIRACL it sets the best score among 0.6B models at 74.11, ahead of 72.20 for \jrrvthree{} and 67.12 for \qwenrerankersmall{}, though it still trails the four times larger \qwenrankerlarge{} at 76.56. RTEB favors compact specialization more strongly. Here \jrrvthreefive{} reaches 70.95, lifting \jrrvthree{} by $2.9$ absolute points and exceeding both the same-size Qwen at 68.41 and the 1.5B mxbai at 70.81. The remaining gap to \qwenrankerlarge{} at 77.68 concentrates on a few legal and medical tasks, as detailed in Table~\ref{tab:rteb}.

\begin{table*}[t]
\centering
\small
\renewcommand{\arraystretch}{1.25}
\setlength{\tabcolsep}{4pt}
\begin{tabular}{@{}l c *{4}{c} @{}}
\toprule
\textbf{Model} & \textbf{\# Param} & \textbf{BEIR} & \textbf{MIRACL} & \textbf{RTEB $^\ddagger$} & \textbf{Struct-IR$^\dagger$} \\
\midrule
\multicolumn{6}{c}{\textit{First-stage retriever}} \\
\midrule
\jinaembedsmall & 0.5B & 56.26 & 65.15 & 64.60 & -- \\
\midrule
\multicolumn{6}{c}{\textit{Second-stage reranker}} \\
\midrule
\qwenrerankersmall & 0.6B & 56.94 & 67.12 & 68.41 & 41.9 \\
\qwenrankerlarge & 4.0B & 62.28 & \textbf{76.56} & \textbf{77.68} & \textbf{55.6} \\
\texttt{mxbai-rerank-base-v2} & 0.5B & 59.58 & 64.90 & 61.44 & 30.4 \\
\texttt{mxbai-rerank-large-v2} & 1.5B & \underline{62.45} & 69.65 & 70.81 & 43.0 \\
\jrrvthree & 0.6B & 62.10 & 72.20 & 68.01 & 38.7 \\
\midrule
\jrrvthreefive & 0.6B & \textbf{63.20} & \underline{74.11} & \underline{70.95} & \underline{48.3} \\
\bottomrule
\end{tabular}
\caption{Overall reranking performance (nDCG@10, \%).
BEIR, MIRACL, and RTEB use top-100 candidates from \jinaembedsmall{}.
Struct-IR$^\dagger$ uses the controlled-pool protocol (Table~\ref{tab:structir}).  $^\ddagger$ The RTEB score excludes the MIRACL average score.
Best per column in \textbf{bold}, second-best \underline{underlined} (first-stage excluded).}
\label{tab:overall}
\end{table*}

\subsection{Multilingual Retrieval}
\label{sec:miracl}

MIRACL covers 18 languages with varying resource levels. \jrrvthreefive{} improves the MIRACL average from $72.20$ to $74.11$ nDCG@10 ($+2.6\%$ relative) over \jrrvthree{}, with the largest absolute gains on Yoruba ($+4.4$), Farsi ($+3.1$), and French ($+3.0$). These lifts track the multilingual expansion beyond the MIRACL~/mMARCO shards of \jrrvthree{} (WebFAQ, SWIMIR, Ruri-v3 Japanese; Section\ref{sec:data}). Indonesian, French, and Spanish remain among the hardest languages by absolute score across models, while Telugu, Thai, and Yoruba score consistently highest. Per-language scores appear in Appendix~\ref{app:miracl}.

\subsection{Domain-Specific Retrieval}
\label{sec:rteb}

RTEB benchmarks rerankers on professional-domain corpora spanning legal case retrieval, financial QA, programming documentation, and medical dialogue, where terminology and relevance conventions differ substantially from web text. Table~\ref{tab:rteb} reports per-task nDCG@10 under the same top-100 protocol as Table~\ref{tab:overall}.

The largest gains over \jrrvthree{} concentrate on the legal case-law tasks that the training mixture was designed to cover. AILA-Statute and AILA-Case improve by $14.0$ and $11.7$ absolute points and recover much of the gap that full-attention \jrrvthree{} left open under citation-dense case law, while on LegalQuAD, where \jrrvthree{} is already strong, the two stay within $1.3$ points. Finance and programming move up more modestly but consistently, reaching $84.85$ on FinBench, $76.57$ on Apps, and $66.20$ on DS1000. On most of these columns \jrrvthreefive{} places second behind either the $4$B \qwenrankerlarge{} or \mxbailarge{}, and it overtakes both on FinQA at $86.91$. Medical lifts are smaller, with ChatDoctor up $1.0$ point and CUREv1 essentially tied with \jrrvthree{}, yet \jrrvthreefive{} still places second on the CUREv1 average and its Spanish and French splits. Aggregated, these per-task gains raise the RTEB macro from $68.01$ to $70.95$, exceeding both the same-size Qwen at $68.41$ and the 1.5B mxbai at $70.81$ while remaining behind the $4$B Qwen baseline at $77.68$. Domain-targeted training at $0.6$B thus closes a substantial fraction of the specialization gap without matching the largest generalist.

\begin{table}[t]
\centering
\small
\renewcommand{\arraystretch}{1.2}
\setlength{\tabcolsep}{5pt}
\caption{Per-task RTEB results (nDCG@10, \%). Columns are models; rows are datasets grouped by domain. Medical CURE reports CUREv1 AVG/en/es/fr.}
\label{tab:rteb}
\resizebox{\textwidth}{!}{%
\begin{tabular}{llccccccc}
\toprule
Domain & Dataset & \shortstack{jina-embed\\v5-s (1st)} & \shortstack{mxbai\\base} & \shortstack{mxbai\\large} & \shortstack{Qwen3\\0.6B} & \shortstack{Qwen3\\4B} & \jrrvthree & \jrrvthreefive \\
\midrule
\multirow{4}{*}{Legal} & AILA-C & 44.77 & 25.91 & \underline{44.77} & 37.31 & \textbf{48.42} & 20.82 & 32.55 \\
 & AILA-S & 53.16 & 34.64 & 43.90 & \underline{88.90} & \textbf{91.06} & 32.15 & 46.16 \\
 & LegalSum & 64.70 & 67.76 & \underline{71.48} & 67.60 & \textbf{72.00} & 69.64 & 70.99 \\
 & LegalQuAD & 63.11 & 46.75 & 78.55 & \underline{88.84} & \textbf{89.27} & 81.84 & 83.09 \\
\midrule
\multirow{3}{*}{Finance} & FinBench & 80.76 & 77.64 & 81.49 & 80.68 & \textbf{88.14} & 83.43 & \underline{84.85} \\
 & HC3Fin & 62.17 & 52.40 & 59.54 & 54.10 & \textbf{68.53} & 62.74 & \underline{64.07} \\
 & FinQA & 55.63 & 82.97 & \underline{86.89} & 81.92 & 85.92 & 85.87 & \textbf{86.91} \\
\midrule
\multirow{6}{*}{Programming} & Apps & 73.22 & 39.68 & 69.30 & 59.94 & \textbf{89.90} & 74.94 & \underline{76.57} \\
 & DS1000 & 61.25 & 62.00 & \textbf{67.59} & 55.43 & 63.30 & 61.94 & \underline{66.20} \\
 & HumanEval & 96.08 & 94.46 & \textbf{99.22} & 93.74 & 97.97 & \underline{98.81} & 98.11 \\
 & MBPP & 90.51 & 89.01 & \textbf{92.88} & 80.27 & 88.59 & 91.64 & \underline{91.84} \\
 & WikiSQL & 93.82 & 99.00 & \textbf{99.25} & 90.77 & 98.51 & 99.02 & \underline{99.14} \\
 & FreshStack & 39.20 & 30.88 & 34.65 & \underline{36.91} & \textbf{46.51} & 25.69 & 31.01 \\
\midrule
\multirow{5}{*}{Medical} & ChatDoctor & 71.10 & 64.53 & \underline{71.52} & 64.41 & \textbf{73.38} & 69.63 & 70.63 \\
 & CURE-AVG & 51.92 & 53.97 & 61.05 & 45.33 & \textbf{63.68} & 62.04 & \underline{62.20} \\
 & CURE-en & 55.71 & 60.02 & 65.56 & 51.81 & \textbf{66.73} & \underline{65.58} & 65.46 \\
 & CURE-es & 50.78 & 51.36 & 58.66 & 42.41 & \textbf{62.16} & 61.05 & \underline{61.12} \\
 & CURE-fr & 49.26 & 50.52 & 58.93 & 41.78 & \textbf{62.16} & 59.48 & \underline{60.01} \\
\bottomrule
\end{tabular}%
}
\end{table}

\subsection{Structured and Field-Constrained Retrieval}
\label{sec:structured}

Structured benchmarks evaluate ranking over product catalogs, bibliographic entities,
and JSON records with heterogeneous schemas. Both benchmarks were designed for
first-stage retrieval, so we adapt them to a reranking setting and are explicit about
the recall regime each measures.

The Struct-IR benchmark of SSRB~\citep{ssrb} indexes millions of semi-structured objects per schema, and a
first-stage retriever alone recovers few gold documents, with in-schema Recall@5 around 0.04 for \jinaembedsmall{}.
End-to-end retrieve-then-rerank is therefore almost entirely recall-bound and separates rerankers poorly.
We instead evaluate under a controlled candidate pool. For each query we inject all gold documents
alongside the 30 hardest first-stage distractors, which guarantees gold presence and measures
field-constrained discrimination rather than retrieval coverage, as reported in Table~\ref{tab:structir}.
This protocol is not comparable to the SSRB retrieval leaderboard.

Under this setting \jrrvthreefive{} reaches macro Recall@5 and nDCG@10 of $23.1$ and $48.3$,
improving over \jrrvthree{} by $5.2$ and $9.6$ absolute points and taking second place across every domain.
Only the $4$B \qwenrankerlarge{} scores higher at $26.4$ and $55.6$, and \jrrvthreefive{} stays well ahead of \mxbailarge{} at $19.5$ and $43.0$
and same-size \qwenrerankersmall{} at $18.3$ and $41.9$.
The largest relative gains over \jrrvthree{} appear on HR, LLM-Agent, and Resume, where constraint-heavy
JSON training most directly transfers.
On Resume, \jrrvthreefive{} ties \qwenrankerlarge{} at the best Recall@5 ($20.5$).

\begin{table*}[t]
\centering
\small
\renewcommand{\arraystretch}{1.25}
\caption{\structir{} controlled-pool reranking (Recall@5 / nDCG@10, \%).
For each query, all gold documents are injected into a pool with the 30 hardest first-stage distractors from \jinaembedsmall{}, isolating reranker discrimination from retrieval recall.}
\label{tab:structir}
\setlength{\tabcolsep}{3.5pt}
\resizebox{\textwidth}{!}{%
\begin{tabular}{lcccccccccccccc}
\toprule
Model & \multicolumn{2}{c}{AVG (macro)} & \multicolumn{2}{c}{Academic} & \multicolumn{2}{c}{Finance} & \multicolumn{2}{c}{HR} & \multicolumn{2}{c}{LLM-Agent} & \multicolumn{2}{c}{Product} & \multicolumn{2}{c}{Resume} \\
\cmidrule(lr){2-3} \cmidrule(lr){4-5} \cmidrule(lr){6-7} \cmidrule(lr){8-9} \cmidrule(lr){10-11} \cmidrule(lr){12-13} \cmidrule(lr){14-15}
 & R@5 & nD@10 & R@5 & nD@10 & R@5 & nD@10 & R@5 & nD@10 & R@5 & nD@10 & R@5 & nD@10 & R@5 & nD@10 \\
\midrule
\texttt{mxbai-rerank-large-v2} & 19.5 & 43.0 & 15.7 & 46.8 & 19.7 & 43.0 & 22.9 & 49.7 & 28.3 & 51.0 & 13.8 & 34.4 & \underline{15.1} & 32.5 \\
\qwenrerankersmall & 18.3 & 41.9 & 15.8 & 48.2 & 17.9 & 40.7 & 22.0 & 47.9 & 24.0 & 45.3 & 14.4 & 35.4 & 14.7 & 33.9 \\
\qwenrankerlarge & \textbf{26.4} & \textbf{55.6} & \textbf{22.6} & \textbf{62.9} & \textbf{27.2} & \textbf{56.3} & \textbf{28.7} & \textbf{60.1} & \textbf{36.6} & \textbf{64.7} & \textbf{21.7} & \textbf{47.1} & \textbf{20.5} & \textbf{42.0} \\
\jrrvthree & 17.9 & 38.7 & 15.8 & 44.0 & 19.6 & 38.2 & 21.1 & 43.7 & 23.6 & 42.7 & 13.7 & 34.4 & 12.5 & 28.7 \\
\midrule
\jrrvthreefive & \underline{23.1} & \underline{48.3} & \underline{18.6} & \underline{52.7} & \underline{22.5} & \underline{44.3} & \underline{26.5} & \underline{56.4} & \underline{34.1} & \underline{58.1} & \underline{14.8} & \underline{36.4} & \textbf{20.5} & \underline{41.3} \\
\bottomrule
\end{tabular}%
}
\end{table*}

STARK~\citep{stark} is a benchmark for retrieval over semi-structured knowledge bases, where the corpus is a knowledge graph of typed entities connected by relations and each query is a natural-language question whose answer is a small set of ground-truth entities.
Following the official protocol, we render each candidate entity together with its relational context and rerank the top-100 candidates from \jinaembedsmall{}, as reported in Table~\ref{tab:stark}.
Because each query has only one or a few correct entities, we report Hit@$k$ and MRR, macro-averaged over Amazon, MAG, and Prime.

On this protocol \jrrvthreefive{} improves over \jrrvthree{} on all three official metrics, raising Hit@1 from $43.4$ to $45.0$, Hit@5 from $63.6$ to $65.4$, and MRR from $53.2$ to $53.8$. It attains the best Hit@5 overall and leads on MAG across the board.
It stays competitive with \mxbailarge{}, which reaches $46.2$ Hit@1 and $53.4$ MRR, while clearly outperforming \qwenrankerlarge{}, \qwenrerankersmall{}, and \mxbaibase{}.

\begin{table*}[t]
\centering
\small
\renewcommand{\arraystretch}{1.25}
\caption{STARK under the official protocol (Hit@1 / Hit@5 / MRR, \%).
Entity attributes are rendered as natural-language prose with relational context; top-100 candidates from \jinaembedsmall{}.}
\label{tab:stark}
\setlength{\tabcolsep}{4pt}
\resizebox{\textwidth}{!}{%
\begin{tabular}{lcccccccccccc}
\toprule
Model & \multicolumn{3}{c}{AVG (macro)} & \multicolumn{3}{c}{Amazon} & \multicolumn{3}{c}{MAG} & \multicolumn{3}{c}{Prime} \\
\cmidrule(lr){2-4} \cmidrule(lr){5-7} \cmidrule(lr){8-10} \cmidrule(lr){11-13}
 & H@1 & H@5 & MRR & H@1 & H@5 & MRR & H@1 & H@5 & MRR & H@1 & H@5 & MRR \\
\midrule
\mxbaibase & 39.0 & 58.7 & 47.8 & 37.0 & 66.7 & 50.1 & 45.2 & 52.4 & 48.7 & 34.7 & 57.1 & 44.4 \\
\mxbailarge & \textbf{46.2} & 62.6 & \underline{53.4} & \textbf{58.0} & 75.3 & \textbf{65.2} & 42.9 & 51.2 & 46.7 & \textbf{37.8} & \textbf{61.2} & \textbf{48.4} \\
\qwenrerankersmall & 30.5 & 51.3 & 40.2 & 40.7 & 67.9 & 53.4 & 34.5 & 45.2 & 39.7 & 16.3 & 40.8 & 27.5 \\
\qwenrankerlarge & 32.5 & 54.1 & 42.1 & 42.0 & 70.4 & 54.5 & 39.3 & 51.2 & 44.2 & 16.3 & 40.8 & 27.5 \\
\jrrvthree & 43.4 & \underline{63.6} & 53.2 & 46.9 & \underline{76.5} & 61.1 & \underline{46.4} & \underline{56.0} & \underline{51.0} & \underline{36.7} & \underline{58.2} & \underline{47.4} \\
\midrule
\jrrvthreefive & \underline{45.0} & \textbf{65.4} & \textbf{53.8} & \underline{48.1} & \textbf{77.8} & \underline{61.6} & \textbf{50.0} & \textbf{57.1} & \textbf{52.8} & \underline{36.7} & \textbf{61.2} & 47.1 \\
\bottomrule
\end{tabular}%
}
\end{table*}

\subsection{Inference Efficiency}
\label{sec:efficiency}

The 3L2G schedule with window $w{=}1024$ reduces per-layer attention cost from $\mathcal{O}(L^2)$ to $\mathcal{O}(L\cdot w)$ on the 17 local layers, so the relative savings should grow with the concatenated listwise sequence length $L$.
We measure wall-clock efficiency on an NVIDIA A100 with batch size~1, top-100 listwise inputs, and FlashAttention-2, contrasting two regimes that share the same ranking depth but differ sharply in token length, as reported in Table~\ref{tab:efficiency}. The short-context regime uses BEIR Natural Questions, with mean query and document lengths of $10.3$ and $145.5$ tokens over $254$ timed queries. The long-context regime uses RTEB AILACasedocs, with mean lengths of $689.8$ and $1{,}904.0$ tokens over $48$ timed queries.

On NQ, \jrrvthreefive{} cuts mean latency from $371$\,ms to $305$\,ms ($1.22\times$ speedup) and raises document throughput from $270$ to $328$\,docs/s.
On AILACasedocs the same model yields a much larger gain, cutting mean latency from $16.1$\,s to $10.3$\,s for a $1.56\times$ speedup and raising prefill throughput from $11.9$k to $18.6$k tokens/s. Hybrid attention thus pays off most when candidate passages are long.
Because production listwise reranking is dominated by a single prefill over a fresh candidate set, these latency reductions directly expand feasible list length and document size at fixed serving budget.

\begin{table}[t]
\centering
\small
\renewcommand{\arraystretch}{1.25}
\caption{Inference efficiency on NVIDIA A100 (batch size~1, top-100 listwise reranking, FlashAttention-2).
NQ is short-context; AILACasedocs is long-context legal text.
Speedup is mean latency of \jrrvthree{} divided by that of \jrrvthreefive{}.}
\label{tab:efficiency}
\setlength{\tabcolsep}{4pt}
\begin{tabular}{@{}l cc cc@{}}
\toprule
 & \multicolumn{2}{c}{NQ (BEIR)} & \multicolumn{2}{c}{AILACasedocs (RTEB)} \\
\cmidrule(lr){2-3}\cmidrule(lr){4-5}
 & \jrrvthree & \jrrvthreefive & \jrrvthree & \jrrvthreefive \\
\midrule
Mean latency (ms) & 371.1 & 305.3 & 16{,}064.9 & 10{,}290.9 \\
Speedup & -- & $1.22\times$ & -- & $1.56\times$ \\
Queries / s & 2.69 & 3.28 & 0.06 & 0.10 \\
Documents / s & 269.5 & 327.6 & 6.22 & 9.72 \\
Prefill tokens / s & 39{,}210 & 48{,}357 & 11{,}884 & 18{,}620 \\
\bottomrule
\end{tabular}
\end{table}

\section{Discussion}
\label{sec:discussion}

\jrrvthreefive{} shows that targeted training can bring a $0.6$B listwise reranker up to par with 4B and larger repurposed LLMs, and even surpass them in some areas. For enterprise retrieval, this argues for investing in small models with focused, competent training rather than defaulting to the largest model available.


Our three-stage training schedule shows that it is possible to get high-quality results from teacher-student model distillation even when there is an architectural mismatch in the attention schedule. Adapting first under the sparse mask, then separately learning to match teacher behavior, is more effective than trying to jointly optimize for both in a single training stage. The training procedure used for \jrrvthreefive{} is specific to the 3L2G architecture, but the outline is general enough to transfer to other attention schedule mismatches.


On BEIR \jrrvthreefive{} mildly improves on both \jrrvthree{} and \qwenrankerlarge{}, and on multilingual MIRACL it shows a similar improvement over \jrrvthree{} while remaining somewhat behind \qwenrankerlarge{}. This reiterates what \jrrvthree{} demonstrated, that listwise rerankers can stay competitive with and even beat much larger rerankers built on other principles, here with roughly one-seventh the parameters of \qwenrankerlarge{}. On domain-specific RTEB and controlled-pool Struct-IR the gains over \jrrvthree{} are very large, but important gaps to \qwenrankerlarge{} remain. RTEB legal and medical tasks, controlled-pool Struct-IR, and low-resource MIRACL languages stay the hardest cases.

On the efficiency side, \jrrvthreefive{} reduces reranking latency across the board, most sharply on long inputs and large candidate lists, since its hybrid schedule lowers per-layer attention cost from $\mathcal{O}(L^2)$ to $\mathcal{O}(L\cdot w)$ on the local layers with $w=1024$. Two limitations remain. Listwise rerankers still carry input constraints that pointwise and late-interaction models avoid, in particular model-fixed upper bounds on candidate count and total candidate length in tokens. We also leave open whether Stage~II can be shortened when the teacher-student attention gap is milder than the full-to-3L2G transition studied here.

\section{Conclusion}
\label{sec:conclusion}

\jrrvthreefive{} makes compact listwise reranking faster and more broadly useful. A single 0.6B model improves inference efficiency, domain coverage, and semi-structured understanding while keeping the cross-document comparison of LBNL interaction and its general-benchmark quality, extending the model class into domains where it previously underperformed. The gains carry a reusable lesson for distillation. When teacher and student differ in attention pattern rather than size, adapting the student under the sparse mask before matching teacher behavior transfers most of the teacher's quality, and the recipe should generalize to other attention-schedule mismatches. We release \jrrvthreefive{}'s weights under a non-commercial license at \url{https://huggingface.co/jinaai/jina-reranker-v3.5}.


\bibliographystyle{plainnat}
\bibliography{references}

\appendix

\section{Detailed BEIR Results}
\label{app:beir}
\begin{table}[t]
\centering
\small
\renewcommand{\arraystretch}{1.25}
\setlength{\tabcolsep}{5pt}
\caption{Reranking performance (nDCG@10, \%) on BEIR with top-100 \jinaembedsmall{} candidates. Columns are models; rows are datasets.}
\label{tab:beir}
\resizebox{\textwidth}{!}{%
\begin{tabular}{lccccccc}
\toprule
Dataset & \shortstack{jina-embed\\v5-s (1st)} & \shortstack{mxbai\\base} & \shortstack{mxbai\\large} & \shortstack{Qwen3\\0.6B} & \shortstack{Qwen3\\4B} & \jrrvthree & \jrrvthreefive \\
\midrule
AVG & 56.26 & 59.58 & \underline{62.45} & 56.94 & 62.28 & 62.10 & \textbf{63.20} \\
TREC-COVID & 79.61 & 85.56 & 83.23 & \underline{88.05} & \textbf{89.66} & 85.33 & 85.29 \\
NFCorpus & 40.09 & 37.68 & 38.43 & \underline{39.07} & \textbf{42.48} & 38.43 & 38.45 \\
NQ & 63.76 & 68.34 & 72.25 & 58.34 & 68.71 & \underline{73.42} & \textbf{73.82} \\
HotpotQA & 69.59 & 80.51 & 81.58 & 77.12 & 80.41 & \underline{82.19} & \textbf{82.41} \\
FiQA & 49.40 & 46.42 & \textbf{51.90} & 42.44 & \underline{51.51} & 48.83 & 50.83 \\
ArguAna & 65.92 & 55.44 & 72.69 & 56.95 & 58.64 & \underline{75.51} & \textbf{77.90} \\
Touche & 31.96 & 30.60 & 31.64 & 32.76 & \textbf{40.30} & 33.40 & \underline{34.14} \\
DBPedia & 44.13 & 49.40 & \underline{51.00} & 44.10 & \textbf{51.37} & 48.59 & 49.57 \\
SCIDOCS & 23.06 & 17.11 & 18.77 & 20.62 & \textbf{25.69} & \underline{22.83} & 22.45 \\
FEVER & 90.36 & 93.28 & \textbf{94.16} & 87.02 & 90.23 & \underline{93.67} & \textbf{94.16} \\
Climate-FEVER & 40.28 & 44.82 & \underline{46.72} & 40.23 & \textbf{47.95} & 37.97 & 44.75 \\
SciFact & 76.99 & 77.20 & \textbf{79.93} & 75.83 & \underline{77.95} & 76.42 & 77.15 \\
Quora & 89.63 & 88.18 & 89.60 & 77.75 & 84.76 & \underline{90.68} & \textbf{90.71} \\
\bottomrule
\end{tabular}%
}
\end{table}

\section{Detailed MIRACL Results}
\label{app:miracl}

\begin{table}[t]
\centering
\small
\renewcommand{\arraystretch}{1.2}
\setlength{\tabcolsep}{5pt}
\caption{Multilingual reranking on MIRACL (nDCG@10, \%). Columns are models; rows are languages.}
\label{tab:miracl}
\resizebox{\textwidth}{!}{%
\begin{tabular}{lccccccc}
\toprule
Language & \shortstack{jina-embed\\v5-s (1st)} & \shortstack{mxbai\\base} & \shortstack{mxbai\\large} & \shortstack{Qwen3\\0.6B} & \shortstack{Qwen3\\4B} & \jrrvthree & \jrrvthreefive \\
\midrule
AVG & 65.15 & 64.90 & 69.65 & 67.12 & \textbf{76.56} & 72.20 & \underline{74.11} \\
AR & 75.19 & 77.59 & 79.90 & 77.77 & \textbf{85.69} & 81.07 & \underline{82.91} \\
BN & 75.81 & 67.85 & 76.47 & 75.00 & \textbf{87.47} & 82.69 & \underline{84.02} \\
DE & 58.74 & 36.42 & 53.65 & 58.33 & \textbf{68.91} & 62.56 & \underline{64.65} \\
EN & 56.94 & 62.56 & 64.13 & 60.99 & \textbf{69.66} & 64.26 & \underline{66.36} \\
ES & 56.57 & 58.04 & 60.95 & 58.84 & \textbf{65.72} & 60.98 & \underline{63.47} \\
FA & 55.66 & 52.04 & 57.82 & 62.58 & \textbf{70.77} & 60.65 & \underline{63.76} \\
FI & 74.03 & 71.95 & 76.96 & 71.68 & \textbf{84.21} & 78.27 & \underline{81.09} \\
FR & 58.07 & 58.19 & 60.71 & 56.46 & \textbf{67.93} & 59.30 & \underline{62.31} \\
HI & 59.72 & 49.89 & 56.72 & \textbf{71.77} & 59.72 & 66.58 & \underline{67.91} \\
ID & 52.73 & 56.08 & 58.45 & 55.44 & \textbf{61.76} & 60.12 & \underline{61.63} \\
JA & 69.12 & 73.01 & 76.18 & 66.84 & \textbf{81.61} & 77.95 & \underline{79.70} \\
KO & 66.18 & 68.10 & 68.33 & 68.65 & \underline{76.51} & 74.67 & \textbf{76.67} \\
RU & 67.56 & 69.27 & 74.44 & 66.27 & \textbf{79.76} & 75.45 & \underline{77.30} \\
SW & 62.25 & 61.38 & 68.04 & 59.53 & \textbf{78.98} & 74.32 & \underline{74.83} \\
TE & 82.02 & 82.62 & \textbf{89.60} & 82.62 & \textbf{89.60} & 87.28 & \underline{87.32} \\
TH & 77.68 & 79.66 & 83.09 & 77.63 & \textbf{88.44} & 84.64 & \underline{85.84} \\
YO & 60.73 & 78.70 & 84.03 & 77.04 & \textbf{89.36} & 80.23 & \underline{84.62} \\
ZH & 63.68 & 62.66 & 64.27 & 60.80 & \textbf{71.97} & 68.53 & \underline{69.61} \\
\bottomrule
\end{tabular}%
}
\end{table}

\section{Model Configuration}
\label{app:config}

\begin{table}[h]
\centering
\caption{Architecture of \jrrvthreefive{} (identical to \jrrvthree{} except attention pattern).}
\begin{tabular}{lc}
\toprule
Parameter & Value \\
\midrule
Backbone & Qwen3-0.6B \\
Layers & 28 (17 local + 11 global under 3L2G) \\
Hidden size & 1024 \\
Attention heads & 16 (GQA, 8 KV heads) \\
Sliding window & 1024 tokens \\
Context length & 131{,}072 tokens \\
Projector & $1024 \rightarrow 512 \rightarrow 512$, ReLU \\
Scoring & cosine similarity in projected space \\
\bottomrule
\end{tabular}
\end{table}

\section{Training Stage Hyperparameters}
\label{app:hyper}

\begin{table}[h]
\centering
\small
\caption{Representative hyperparameters by training stage.}
\begin{tabular}{lccc}
\toprule
 & Stage I & Stage II & Stage III \\
\midrule
Attention & full & 3L2G SWA & 3L2G SWA \\
Trainable & all & attn-only, then all/LoRA & all + distill head \\
Learning rate & $6\times10^{-6}$ & $5\times10^{-5}\rightarrow 6\times10^{-6}$ & $5\times10^{-5}$ \\
Negatives & 45 & 45 & 30 \\
Temperature $\tau$ & 0.05 & 0.05 & 0.25 (distill) \\
Max steps & 25k & 15k + 15k & 5k--25k \\
$\beta_{1}$ (score-level MSE) & -- & -- & 0.65 \\
$\beta_{2}$ (state-level MSE) & -- & -- & 0.65 \\
$\beta_{3}$ (embedding cosine distance) & -- & -- & 0.65 \\
$\alpha_{1}$ (in-context similarity) & -- & -- & 0.65 \\
$\alpha_{2}$ (embedding dispersion) & -- & -- & 0.25 \\
\bottomrule
\end{tabular}
\end{table}

\end{document}